\documentclass[%
reprint,
superscriptaddress,
amsmath,amssymb,
aps,
prl,
floatfix,
nofootinbib
]{revtex4-1}

\usepackage{graphicx}
\usepackage{color}
\usepackage{subcaption}
\def\ch{{\rm ch}}
\def\cs{{\rm c}}
\def\td{{\rm td}}
\def\sp{{\rm sp}}

\DeclareMathOperator*{\argmax}{arg\,max}
\DeclareMathOperator{\arccosh}{arccosh}

\begin{document}

\title{Two-Scale Oscillons}
\author{Chang Liu}
\email{cliu712@aucklanduni.ac.nz}
\affiliation{Department of Physics, The University of Auckland, Private Bag 92019, Auckland, New Zealand}
\author{Richard Easther}
\email{r.easther@auckland.ac.nz}
\affiliation{Department of Physics, The University of Auckland, Private Bag 92019, Auckland, New Zealand}

\date{\today}

\begin{abstract} \noindent
Oscillons are spatially stationary, quasi-periodic solutions of nonlinear field theories seen in settings ranging from granular systems, low temperature condensates and  early universe cosmology. We describe a new class of oscillon in which the spatial envelope can have ``off centre'' maxima and pulsate on timescales much longer than the fundamental frequency. These  are exact solutions of the 1-D sine-Gordon equation and we demonstrate numerically that similar solutions exist in up to three dimensions for a range of potentials. The dynamics of these solutions match key properties of  oscillons that may form after cosmological inflation in string-motivated monodromy scenarios.
\end{abstract}

\maketitle

\noindent
Oscillons are localized, oscillatory solutions to non-linear wave equations with  lifetimes far longer than the  fundamental frequency of the underlying system. Oscillon-like solutions have been observed in many systems, including Bose-Einstein condensates \cite{PhysRevA.91.023631}, nonlinear optics \cite{PhysRevLett.108.093901}, vibrated granular materials \cite{Copeland:1995fq,Tsimring:1997zz}, while the  early universe may contain a phase during which most mass-energy is localized in oscillons   \cite{Copeland:1995fq,Hindmarsh:2006ur,Hindmarsh:2007jb,Amin:2010dc,Amin:2010xe,Gleiser:2011xj,Amin:2011hj,Zhou:2013tsa,Amin:2013ika}. Beyond their importance to a broad range of experimental and theoretical domains, oscillons are fascinating  nonlinear systems as their longevity is a product of their  dynamics \cite{Copeland:1995fq,PhysRevD.80.125037} rather than  symmetries of the underlying systems.     

%Previous studies have identified Gaussian \cite{Copeland:1995fq,PhysRevD.80.125037,Gleiser:2008ty} and flat-top \cite{Amin:2010jq} oscillons, and the sine-Gordon breather  \cite{:/content/aip/journal/jmp/25/7/10.1063/1.526415}. 

This Letter describes a new category of oscillons, whose  rich phenomenology is characterized by long term pulsations superimposed on the fundamental oscillatory dynamics and spatial profiles with  ``off-centre'' peaks. These solutions share  hitherto unexplored properties of oscillons seen in simulations of the primordial universe following axion monodromy inflation \cite{stringInflationBook,Silverstein:2008sg,McAllister:2008hb,Flauger:2009ab,Amin:2010dc, Amin:2011hj}. We  construct these models as exact solutions to the sine-Gordon equation in one dimension and   numerically explore similar solutions in the sine-Gordon, $\phi^6$ and axion-monodromy potentials in up to three spatial dimensions. 
 
 \medbreak\noindent {\em Exact Solutions:} Consider the Klein-Gordon equation
 \begin{equation}\label{kg}
 \Box u(t,\bar{x}) + \frac{ d V(u)}{du} = 0
 \end{equation}
where  $\Box$ is the d'Alembertian.  With the sine-Gordon potential $V(u) = 1-\cos(u)$,  this system is integrable in $1+1$D Minkowski space and has  many exact solutions,  including the well-known ``breather'' 
\begin{equation}\label{single}
  u(t,x) = 4 \arctan\left[ \frac{\sqrt{1-\omega^2} \cos \omega t}{\omega \cosh \sqrt{1-\omega^2} x} \right] \, .
\end{equation}
When the characteristic frequency $\omega\to1$, the argument of the $\arctan$ is small, and eq.~(\ref{single}) reduces to a Gaussian  
\begin{equation}
  u(t,x) \approx \frac{4\sqrt{1-\omega^2}\cos\omega t}{\cosh \sqrt{1-\omega^2} x} \approx A\exp\left(-\frac{x^2}{R^2}\right)\cos\omega t \,
\end{equation}
where the amplitude $A = 4\sqrt{1-\omega^2}$ and the width at half maximum $R=(\arccosh 2) /\sqrt{(\log 2)(1 - \omega^2)}$. One may use the Gaussian oscillon as a template, finding stability and existence conditions as a function of $A$ and $R$  \cite{Gleiser:2008ty}. Separately, ``flat-top" solutions are found with a sixth-order potential \cite{Amin:2010jq}, along with rotationally symmetric generalizations beyond one dimension. 

Motivated by the complex dynamics seen in the oscillon dominated phase \cite{Amin:2011hj} following monodromy inflation \cite{Silverstein:2008sg,McAllister:2008hb,Flauger:2009ab} and similar solutions noted in Refs.~\cite{Hindmarsh:2006ur,Amin:2010xe,Salmi:2012ta} we  construct an exact  ``double-breather'' in the 1D sine-Gordon model, 
 \begin{eqnarray}\label{twoscale1}
  u(t,x)&=&4 \arctan \frac{(\omega_1^2-\omega_2^2)(\ch_1\,\cs_2+\ch_2\,\cs_1)}{\td(t)+\sp(x)} \, , \\
    \td(t) &=& (\omega_1^2+\omega_2^2)\,\cs_1\cs_2 +2 \sin \omega_1 t \sin \omega_2 t \, , \\
  \sp(x) &=& (\mu_1^2+\mu_2^2)\,\ch_1\ch_2 - 2  \sinh \mu_1 x \sinh \mu_2 x  \, .
\end{eqnarray}
\begin{equation}\label{twoscale2}
    \cs_i = \frac{\cos \omega_i t}{\omega_i} \quad \ch_i = \frac{\cosh \mu_i x}{\mu_i}  \quad   \mu_i=\sqrt{1-\omega_i^2} \, .
\end{equation}    
This ``non-linear superposition'' of two simple breathers can be constructed via the B\"acklund transformation  \cite{Dodd499, hietarinta1997introduction, Cuenda20111047} or the  inverse scattering transform \cite{ablowitz, :/content/aip/journal/jmp/51/12/10.1063/1.3520596}, and may be verified by direct calculation. The solution is antisymmetric with respect to  $\omega_1$ and $\omega_2$ and $0<\omega_i<1$; when either $\omega_i \rightarrow 1$ we recover the single-frequency breather.

\begin{figure}
\includegraphics[width=\linewidth]{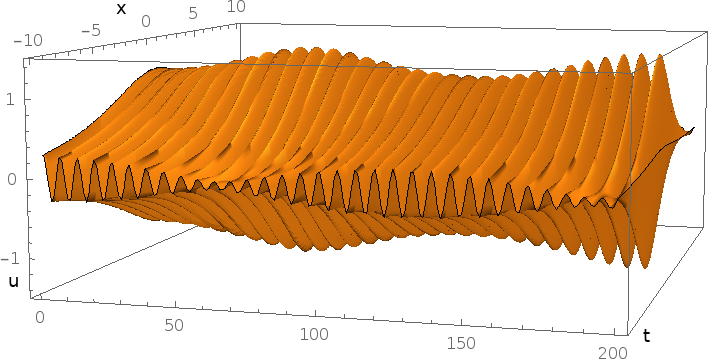} \\
\includegraphics[width=\linewidth]{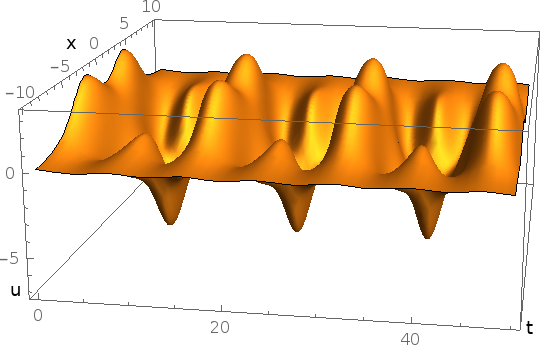}
 \caption{[Top]  Small amplitude  ($\omega_1=0.999$, $\omega_2=0.95$) resembling a simple oscillon with a slow, superimposed pulsation. [Bottom] Large-amplitude oscillon  ($\omega_1=0.8$, $\omega_2=0.4$) with a clear double peak at around $x=\pm3$.\label{profiles}}
\end{figure}

\medbreak\noindent {\em Qualitative Description:}    Fig.~\ref{profiles} shows two typical field profiles; further examples and animations are included in the Supplementary Material. Two key features of these solutions are long-period pulsations superimposed on the oscillon dynamics and spatial profiles with peaks displaced from the origin.    

\begin{figure}
  \centering
  \includegraphics[width=0.45\textwidth]{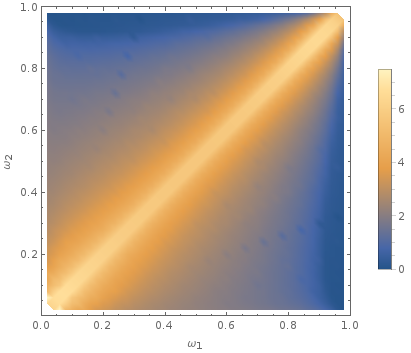}\vskip10pt
  \includegraphics[width=0.45\textwidth]{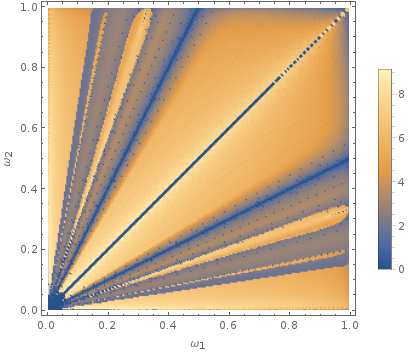}
  \caption{Values $\langle r_{\rm max}\rangle$ (upper) and  $\log{S}$ (lower) for 1D sine-Gordon.}\label{sine-1d}
\end{figure}

We quantify the peak-displacement via
\begin{equation}
  \langle r_{\rm max} \rangle  = \left\langle \argmax_{x < r_0} |u(t,x)| \right\rangle_t
\end{equation}
the time-average of the peak field value. Spurious maxima associated with outgoing radiation can appear when $u$ is close to zero for moderate $r$; these are excluded via the cutoff  $r_0$. We quantify the strength of pulsations of the envelope via $S$, the product of the period and amplitude of the dominant periodic mode in the time series of maxima at the origin, $|u(t,0)|$.  Plots of $S$ and $r_{\rm max}$ for the exact ``double breather'' are shown in Fig.~\ref{sine-1d}. If one of the $\omega_i$ is close to unity $r_{\rm max}$ is zero. 

%In the limit  $\delta=\omega_1-\omega_2$ is small the peak displacement at $t=0$ can be shown to be
%\begin{equation}\label{disp}
%  x_d = \frac{1}{2\mu}\cosh^{-1}\left(\frac{2\mu^4}{3\omega^4}\left(1-\frac{2\omega}{\delta}+\frac{4\omega^2}{\delta^2}\right)-1\right)
%\end{equation}
%where $\omega=\omega_1$ and $\mu=\sqrt{1-\omega^2}$.   The  double peak grows more prominent if $\omega_1 \rightarrow  \omega_2$, accounting for the large values of  $r_{\rm max}$  near the diagonal in Fig.~\ref{sine-1d}; conversely if $x_d$ is imaginary  $\langle r_{\rm max}\rangle=0$.

\begin{figure*}
  \centering
    \includegraphics[width=0.32\textwidth]{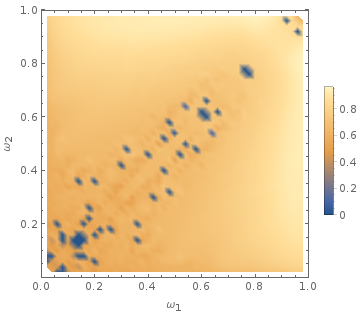}
    \includegraphics[width=0.32\textwidth]{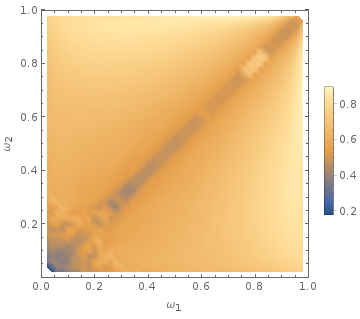}
    \includegraphics[width=0.32\textwidth]{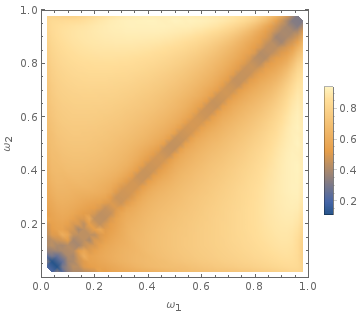} \\\vskip-5pt
    \includegraphics[width=0.32\textwidth]{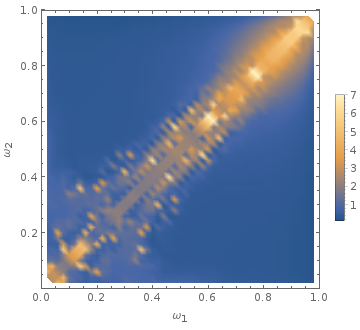}
    \includegraphics[width=0.32\textwidth]{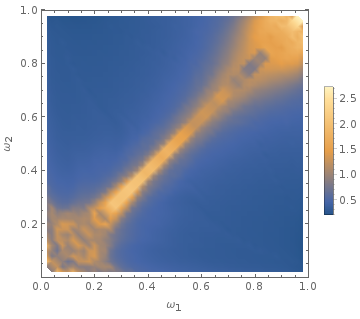}
    \includegraphics[width=0.32\textwidth]{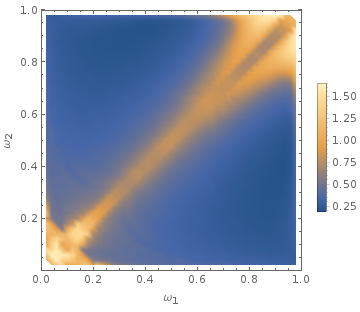} \\\vskip-5pt
    \includegraphics[width=0.32\textwidth]{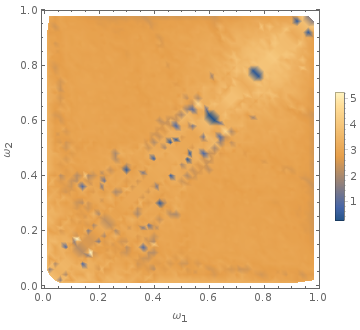}
    \includegraphics[width=0.32\textwidth]{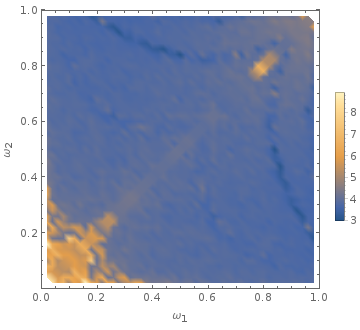}
    \includegraphics[width=0.32\textwidth]{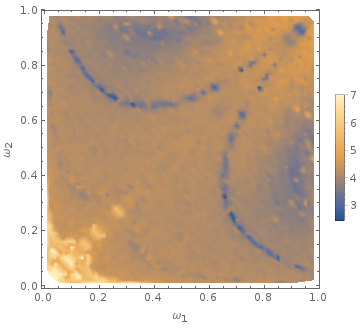}
    \caption{Qualitative properties of pulsating oscillon solutions for the axion monodromy potential. Top row: $E/E_{\rm init}$ at $t=4000$.
      Middle row: $\langle r_{\rm max}\rangle$.
      Bottom row: Strength of slow modes, $\log{S}$.\quad
      Left to right: 1D, 2D, 3D. We see that 50\% of the initial energy typically remains in the oscillon, and the peak is displaced furthest from the origin when $\omega_1 \approx \omega_2$. All choices of  $\omega_1$ and $\omega_2$ lead to a pulsating envelope, which is most pronounced when $\omega_1$ and $\omega_2$ are small in 2D and 3D. The ``holes'' in the 1D solution are associated with cases where the initial profile splits into two single oscillons moving away from the origin.}\label{axion-monodromy}
\end{figure*}

\medbreak\noindent{\em General Potentials and Dimensionality:}   We  assess the existence of qualitatively similar solutions to eqs.~\ref{twoscale1}--\ref{twoscale2}  for the axion monodromy  and sixth-order potentials in 1, 2 and 3 dimensions, and the sine-Gordon in 2 and 3 dimensions. The relevant potentials are:
\begin{eqnarray}
  V(u) &=& \sqrt{1+u^2}-1 \label{axion-potential} \, ,\\
  V(u) &=& \frac{u^2}{2} - \frac{u^4}{4!} +g \frac{ u^6}{6!} \, .
\end{eqnarray}
Note that possible constant factors are set to unity by coordinate transformations and/or rescaling $u$ \cite{Amin:2010dc,Amin:2011hj}.

We solve the Klein-Gordon equation numerically with an added ``viscosity''  at large $r$ to absorb outgoing radiation. We assume a radially symmetric field configuration with the exact 1D sine-Gordon solution as the initial profile, in contrast to  a Gaussian ansatz of e.g. Refs.~\cite{Hindmarsh:2006ur,Gleiser:2008ty,Gleiser:2009ys,Salmi:2012ta}.  For the numerical solutions we quantify the overlap with the exact double-breather profile via the fraction of the initial  energy contained within a fixed radius at late times, recalling that the energy density is:
\begin{equation}
  e(t,\bar{x}) = \frac{1}{2}\left(u_t^2+ \nabla u\cdot\nabla u \right) + V(u) \, .
\end{equation}
Fig.~\ref{axion-monodromy} shows our qualitative measures for the axion monodromy potential (eq.~\ref{axion-potential}) in 1, 2, and 3D. In almost all cases we see an oscillon with $>50\%$ of the initial energy; $\langle r_{\rm max}\rangle$ has a similar dependence on $(\omega_1,\omega_2)$ to the 1D sine-Gordon case and we see significant pulsations across the $(\omega_1,\omega_2)$ plane.

For the sine-Gordon potential, we find long-lived, pulsating oscillons in 2D (in agreement with  \cite{Hindmarsh:2006ur}) but $E/E_{\rm init}$ is much less than unity, suggesting these oscillons are  less closely related to their exact 1D analogues. In 3D, our initial ansatz does not yield stable oscillons over most of the $(\omega_1,\omega_2)$ plane. The sixth-order potential  supports pulsating oscillons  in 1, 2 and 3D when $g$ is small; for large $g$ they only exist if  $\omega_1$ and $\omega_2$ are close to unity. 

\begin{figure}
  \centering
  \includegraphics[width=0.45\textwidth]{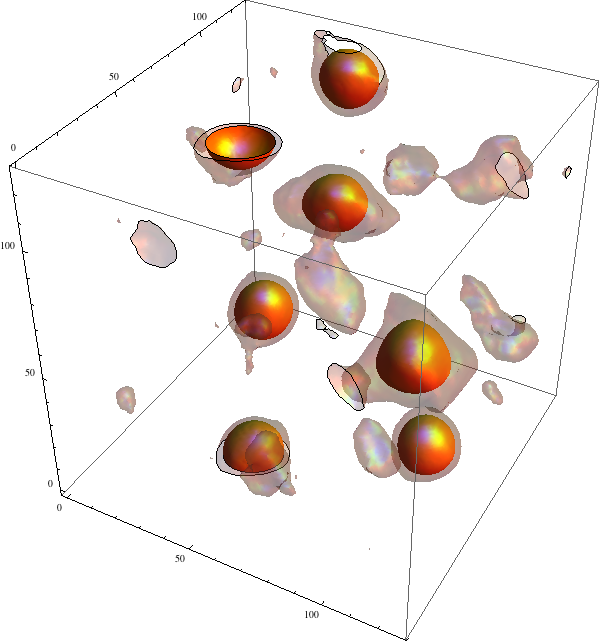}
  \caption{Oscillons following monodromy inflation  \cite{Easther:2010qz}. Contours show regions of mean and 4 $\times$ mean density. }\label{oscillons}
\end{figure}

\begin{figure}
  \centering
  \includegraphics[width=0.45\textwidth]{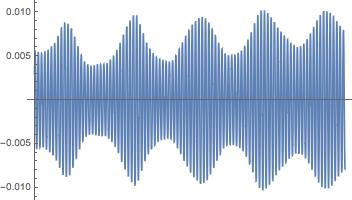} 
    \caption{Central value of an oscillon extracted from full 3D numerical solution.}
  \label{raw}
\end{figure}

\begin{figure}
  \centering
  \includegraphics[width=0.45\textwidth]{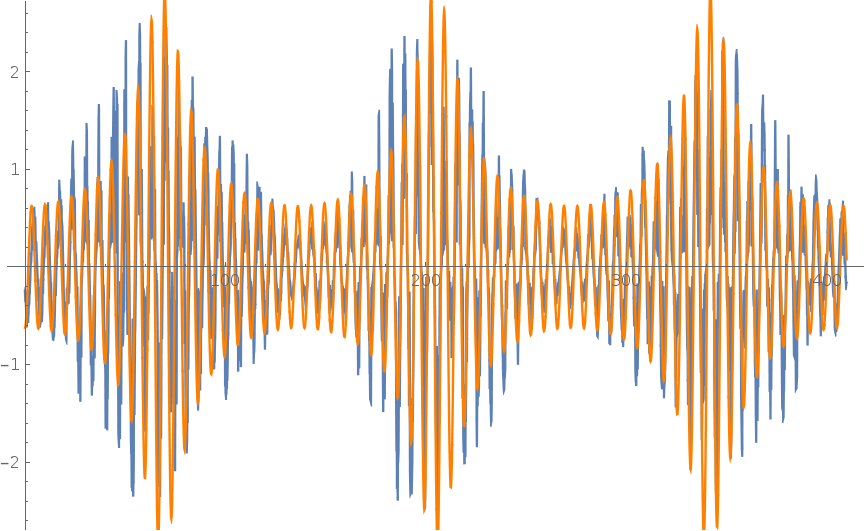}
  \caption{[Blue] Field value at the center of an oscillon from 3D monodromy simulation. [Orange] Fitted profile eq.~(\ref{fitprof}); $A=0.863$, $\omega=0.946$, $\epsilon=0.91$,  $\delta=0.046$.}\label{simul-prof}
\end{figure}

\begin{figure}
  \centering
  \includegraphics[width=0.45\textwidth]{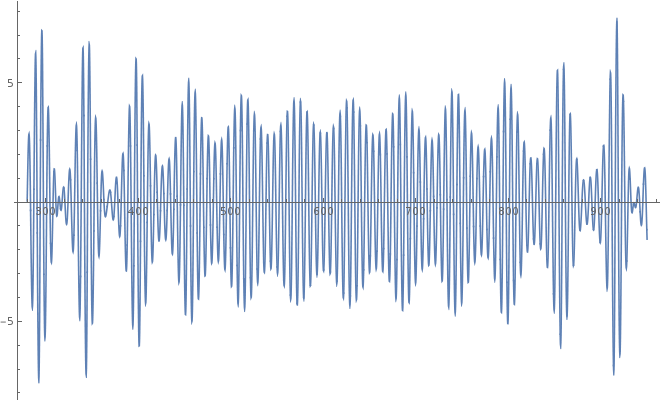}
  \includegraphics[width=0.45\textwidth]{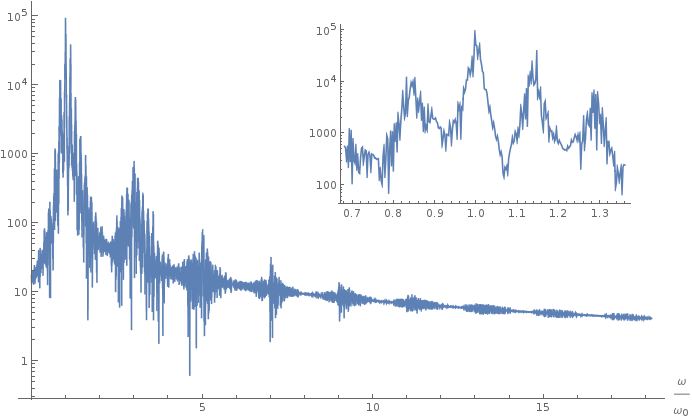}
  \caption{[Top] Spherically symmetric 3D monodromy oscillon initialised with 1D
    sine-Gordon profile $\omega_1=0.999$, $\omega_2=0.950$.  [Bottom] Fourier spectrum. Inset shows  peaks around the dominant mode $\omega_0\approx0.95$.}\label{fourier}
\end{figure}
 
%% \begin{figure}
%%   \centering
%%   \includegraphics[width=0.45\textwidth]{analy-profile.png}
%%   \caption{Exact solution with $\omega_1=0.98$, $\omega_2=0.95$.}\label{analy-prof}
%% \end{figure}

\medbreak\noindent{\em  Axion-monodromy Inflation:} In  axion-monodromy inflation  \cite{Silverstein:2008sg,McAllister:2008hb,Flauger:2009ab} parametric resonance can be followed by an oscillon-dominated phase  \cite{Amin:2011hj,Zhou:2013tsa}. A  representative configuration  extracted from a simulation of the expanding, post-inflationary universe \cite{Easther:2010qz} is shown in Fig.~\ref{oscillons}. These monodromy oscillons can exhibit strong pulsations with  complex dynamics, as illustrated in Fig.~\ref{raw}.  One might wonder if some of the longest-period modulations are induced by the oscillon's interaction with its environment,  finite resolution effects in a 3D code, or the overall expansion of the universe, a necessary  ingredient for the resonant production of oscillons in the cosmological simulations. However, similar oscillon solutions are seen in the spherically symmetric code initialised with a 1D sine-Gordon profile with one of the $\omega_i\to1$. Ref.~\cite{Salmi:2012ta} previously investigated 3D oscillons in a monodromy-like potential initialised with a Gaussian profile. For these solutions the Fourier spectrum of $u(t,0)$ has several distinct peaks, but the dominant mode has $\sim 100\times$ more power than the next highest peaks, but the corresponding spectrum seen here is far more complex,  as can be seen from  Fig.~\ref{fourier}. The overall energy of these oscillons can decrease abruptly, suggesting that on long timescales the solutions are aperiodic. The oscillons seen in the full 3D simulations (Fig.~\ref{oscillons}) do not have off-center peaks.  

In the limit where  both $\omega$ terms are  close to unity, using trigonometric identities to combine the $\cosh$ and $\cos$ terms the solution in eqs.~\ref{twoscale1}--\ref{twoscale2} reduces to
\begin{equation} \label{fitprof}
  u(t,r) = A\cos\omega t \frac{\sqrt{1+\epsilon \cos \delta t}}{\cosh(r/R) + \epsilon \cos \delta t}
\end{equation}
where $\epsilon<1$ and $\delta = \omega_1-\omega_2 \ll 1$. This provides an analytical template for further exploration of these solutions and yields a tolerable fit for many (although not all) of the oscillons found in the 3D solutions on moderate timescales, as seen in Fig.~\ref{simul-prof}. 

\medbreak\noindent{\em Conclusion:} We have explored  pulsating oscillons starting from exact solutions to the 1D sine-Gordon equation. They are characterised by a long-term pulsation of their envelope and spatial profiles with peaks displaced from the origin. We find qualitatively similar solutions in the  2D sine-Gordon model, and the $\phi^6$ and axion-monodromy potentials in 1, 2, and 3D. These solutions illuminate the complex properties of  oscillons that can form in the very early universe \cite{Amin:2011hj}, point to rich new possibilities for  laboratory  studies of nonlinear phenomena, and reveal a rich new class of solutions to the nonlinear wave equation.

\medbreak\noindent{\em Acknowledgements}  We thank Mustafa Amin, Ed Copeland, Miro Erkintalo, Mark Hindmarsh, Eugene Lim, Joshua Rippon, and Bonnie Yu for useful discussions. In particular, we are grateful to Amin and Rippon for creating the numerical code used in part of this work.  We acknowledge the use of the New Zealand eScience Infrastructure (NeSI) high-performance computing facilities, which are funded jointly by NeSI's collaborator institutions and through the Ministry of Business, Innovation \& Employment's Research Infrastructure programme [\url{http://www.nesi.org.nz}]


%\bibliographystyle{apsrev4-1}
%\bibliography{prl-submissionReferences}

\end{document}